\documentclass{emulateapj}

\newcommand{\Halpha}{H$\alpha$}
\newcommand{\halpha}{H$\alpha$}
\newcommand{\target}{V404~Cyg}
\newcommand{\chandra}{{\it Chandra}}
\slugcomment{Submitted to the Astrophysical Journal Letters}

\shorttitle{X-ray and Optical Variability in V404 Cyg}
\shortauthors{Hynes et al.}

\begin{document}


\title{Correlated X-ray and Optical Variability in V404~Cyg in Quiescence}

\author{R. I. Hynes\altaffilmark{1,2},
	P. A. Charles\altaffilmark{3},
        M. R. Garcia\altaffilmark{4},
	E. L. Robinson\altaffilmark{1},
	J. Casares\altaffilmark{5},
	C. A. Haswell\altaffilmark{6},
	A. K. H. Kong\altaffilmark{4},
	M. Rupen\altaffilmark{7},
	R. P. Fender\altaffilmark{8},
	R. M. Wagner\altaffilmark{9},
	E. Gallo\altaffilmark{8},
        B. A. C. Eves\altaffilmark{6},
	T. Shahbaz\altaffilmark{5},
	C. Zurita\altaffilmark{10}
}

\altaffiltext{1}{McDonald Observatory and Department of Astronomy, 
The University of Texas at Austin, 
1 University Station C1400, Austin, Texas 78712, USA}
\altaffiltext{2}{Hubble Fellow; rih@astro.as.utexas.edu}
\altaffiltext{3}{School of Physics and Astronomy, 
The University of Southampton, Southampton, SO17 1BJ, UK}
\altaffiltext{4}{Harvard-Smithsonian Center for Astrophysics, 
60 Garden Street, MS-67, Cambridge, MA 02138, USA}
\altaffiltext{5}{Instituto de Astrof\'\i{}sica de Canarias, 38200 La Laguna,
Tenerife, Spain}
\altaffiltext{6}{Department of Physics and Astronomy, The Open 
University, Walton Hall, Milton Keynes, MK7 6AA, UK}
\altaffiltext{7}{National Radio Astronomy Observatory, Array
  Operations Center, 1003 Lopezville Road, Socorro, NM 87801}
\altaffiltext{8}{Astronomical Institute `Anton Pannekoek', University 
of Amsterdam, Kruislaan 403, 1098 SJ Amsterdam, the Netherlands}
\altaffiltext{9}{Large Binocular Telescope Observatory, University of 
Arizona, 933 North Cherry Avenue, Tucson, AZ 85721}
\altaffiltext{10}{Centro de Astronomia e Astrof\'\i{}sica de Universidade
  de Lisboa, Observat\'{o}rio Astron\'{o}mico de Lisboa, Tapada da Ajuda, 
1349-018 Lisboa, Portugal}

\begin{abstract}
We report simultaneous X-ray and optical observations of \target\ in
quiescence.  The X-ray flux varied dramatically by a factor of $\ga
20$ during a 60\,ks observation.  X-ray variations were well
correlated with those in \halpha, although the latter include an
approximately constant component as well.  Correlations can also be
seen with the optical continuum, although these are less clear.  We
see no large lag between X-ray and optical line variations; this
implies they are causally connected on short timescales.  As in
previous observations, \halpha\ flares exhibit a double-peaked profile
suggesting emission distributed across the accretion disk.  The peak
separation is consistent with material extending outwards to at least
the circularization radius.  The prompt response in the entire
\halpha\ line confirms that the variability is powered by X-ray
(and/or EUV) irradiation.
\end{abstract}


\section{Introduction}
Accretion onto black holes is observed over a wide range of
luminosities.  While the upper end of the luminosity range, including
bright X-ray binaries and active galactic nuclei, is relatively
accessible, quiescent or near-quiescent accretion is more difficult to
study.  There remain large uncertainties about the structure of the
accretion flow in quiescence (see e.g., \citealt{Narayan:2002a}), and
it is possible that the energy output could be dominated by a jet
rather than by the accretion flow itself \citep{Fender:2003a}.  These
accreting black holes emit across the electromagnetic spectrum from
radio to X-rays, so multiwavelength studies can be used to disentangle
different sources from different regions of the inflow, outflow, or
jet, and to establish causal connections between them.

\begin{figure*}
\includegraphics[angle=90,scale=0.7]{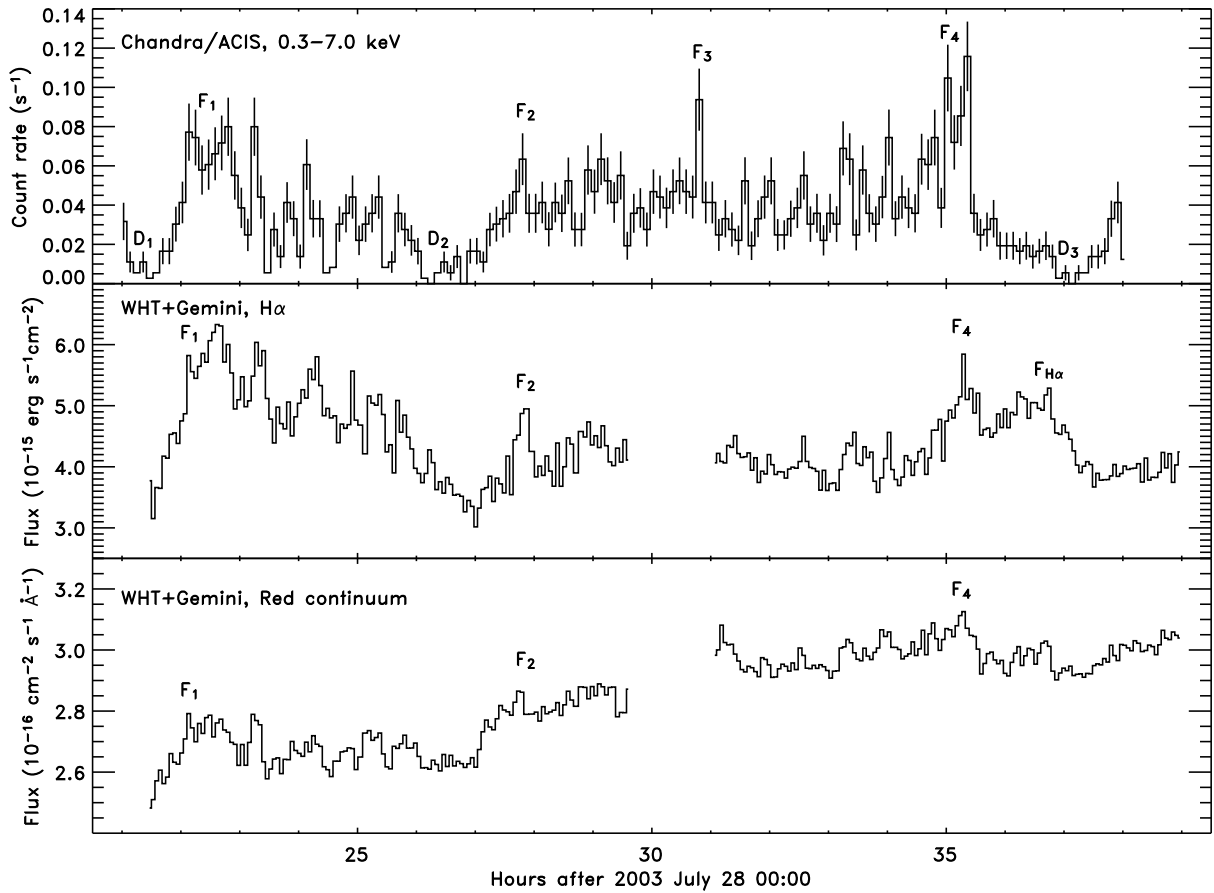}
\caption{Simultaneous X-ray and optical lightcurves.  Annotations
  refer to pronounced dips in the X-ray lightcurve (D$_1$--D$_3$),
  selected flares (F$_1$--F$_4$), and a flare seen in \halpha\ that is
  not present in X-rays (X).  X-ray data is plotted in 400\,s bins,
  and approximately 1\,cnt\,s$^{-1}$ corresponds to an unabsorbed
  0.3--7.0\,keV luminosity of $3.2\times10^{34}$\,erg\,s$^{-1}$.  
  The first half of the optical coverage is from the WHT, the second
  half from Gemini.  WHT
  corresponds to individual 200\,s exposures, and Gemini shows 40\,s
  exposures binned by a factor of four to match the WHT resolution.
  The continuum bandpass is 6300--6500\,\AA\ plus 6620--6820\,\AA.
  The overall continuum rise is the ellipsoidal modulation of the
  companion star.  Error-bars are omitted from the optical data for
  clarity, but are much smaller than those on the X-ray data.}
\label{MWLCFig}
\end{figure*}

This approach has had very little application for quiescent systems to
date.  Of the stellar mass black hole population, the most accessible
quiescent object is V404~Cyg.  V404~Cyg is known to vary in X-rays
(\citealt{Wagner:1994a}; \citealt{Kong:2002a}), optical
(\citealt{Wagner:1992a}; \citealt{Casares:1993a};
\citealt{Pavlenko:1996a}; \citealt{Hynes:2002a};
\citealt{Zurita:2003a}; \citealt{Shahbaz:2004a}), IR
\citep{Sanwal:1996a}, and radio \citep{Hjellming:2000a}, but none of
these studies were coordinated.  \citet{Hynes:2002a} established that
optical emission line variations are correlated with the optical
continuum.  They also found that the emission line flares exhibited a
double-peaked line profile, suggestive of emission distributed across
the accretion disk (see e.g., \citealt{Horne:1986a}) rather than
arising in localized regions.  This was attributed to irradiation of
the outer disk by the variable X-ray source, and hence it was
predicted that the X-ray variations should be correlated with the
optical.  Such correlated variability, also attributed to irradiation,
is commonly seen in X-ray bright states in both neutron star systems
and black holes (e.g.\ \citealt{Grindlay:1978a};
\citealt{Petro:1981a}; \citealt{Hynes:1998a}; and many other works),
but had not been directly observed in quiescent systems.  It is also
usually only detected in the optical continuum yielding no kinematic
information.

Here we report initial results from a coordinated, multiwavelength
campaign to test this prediction.  This included X-ray, near-UV,
optical, and radio coverage, but this letter discusses only the
results from comparing X-ray data with optical spectroscopy.  Future
works will study the variability properties in more detail and examine
the broad-band spectral energy distribution.

\section{Observations}
\subsection{X-ray data}
\chandra\ observations on 2003 July 28/29 used the ACIS camera, in a
single 61.2\,ks observation spanning binary phases 0.51--0.62.  The
source was positioned on the ACIS-S3 chip and the $1/8$ sub-array mode
was used to reduce the frame-time to 0.4\,s and hence ensure that
pile-up was negligible.  Data analysis used Ciao 3.0.  Source events
were extracted from a 3\arcsec\ radius aperture, retaining events with
energies of 0.3--7.0\,keV; a total of 1941 such events were recorded
corresponding to an average count rate of 0.03\,cnt\,s$^{-1}$, a
factor of five lower than the previous observation
(\citealt{Garcia:2001a}; \citealt{Kong:2002a}).  The background count
rate was approximately constant, produced $\sim 4$ counts in the
source aperture, and was neglected for subsequent analysis.  The
spectrum will not be discussed here, but was very similar to that seen
by \citet{Kong:2002a}.  Based on this spectrum, and assuming a
distance of 3.5\,kpc, we estimate that 1\,cnt\,s$^{-1}$ corresponds to
an unabsorbed 0.3--7.0\,keV luminosity of
$3.2\times10^{34}$\,erg\,s$^{-1}$.

\subsection{Optical spectrophotometry}
The first half of our time-resolved optical spectrophotometry was
obtained with the ISIS dual-arm spectrograph on the WHT.  To maximize
efficiency and minimize readout time and noise, we used the single
red-arm mode with the R316R grating and MARCONI2 CCD.  Exposure times
were 200\,s, with $\sim17$\,s dead-time between exposures.  A
4\arcsec\ slit was used to maximize photometric accuracy, so our
spectral resolution was determined by the seeing (median
$\sim1.3$\arcsec), and was typically $\sim5.5$\,\AA\
(250\,km\,s$^{-1}$).  Bias correction and flat fielding were performed
using standard IRAF\footnote{IRAF is distributed by the National
Optical Astronomy Observatories, which are operated by the Association
of Universities for Research in Astronomy, Inc., under cooperative
agreement with the National Science Foundation.} techniques.  The slit
was aligned to cover the same comparison star as used for our previous
observations of the target, and spectra of both of these stars, and
the nearby blended star, were extracted with the same techniques
previously described (\citealt{Hynes:2002b}; \citealt{Hynes:2002a}).
Wavelength calibration was performed relative to a single observation
of a CuNe/CuAr lamp.  Time-dependent variations in the wavelength
calibration were corrected using Telluric absorption features.  The
on-slit comparison star was calibrated relative to Kopff 27
\citep{Stone:1977a}, and all spectra of \target\ were calibrated
relative to this on-slit comparison.

The second half of our optical coverage was provided by the GMOS
spectrograph on Gemini-N.  We used the R831 grating and standard EEV
CCDs.  Exposure times were 40\,s and binning and windowing were used
to reduce the dead-time between exposures to 12\,s.  A 5\arcsec\ slit
with 1.1\arcsec\ median seeing resulted in a spectral resolution of
5.0\,\AA\ (230\,km\,s$^{-1}$).  Data reduction, spectral extraction,
and wavelength and flux calibration were performed in the same way as
for the WHT data.  Wavelength calibration used a CuAr lamp, and flux
calibration was performed relative to the same on-slit comparison star
as used for the WHT observations.

\section{Lightcurves}
X-ray and optical lightcurves are shown in Fig.~\ref{MWLCFig}.
Dramatic X-ray variability is clearly present with a dynamic range of
greater than a factor of twenty, comparable to that seen by
\citet{Wagner:1994a} on longer timescales.  The lowest count rates
seen in three dips (D$_1$--D$_3$) correspond to a luminosity of
$\la10^{32}$\,erg\,s$^{-1}$, which is comparable to the X-ray
luminosities of other quiescent black holes \citep{Garcia:2001a}.
Clear correlations are seen between the X-ray flux and that in both
\halpha\ and the optical continuum in overall trends and in the
distinct flares (e.g., F$_1$, F$_2$, and F$_4$).  One unusually fast
X-ray flare occurred (F$_3$) lasting for $\la 200$\,s and reaching a
peak count rate (unresolved in Fig.~\ref{MWLCFig}) in excess of
0.25\,cnt\,s$^{-1}$.  While \halpha\ generally tracks the X-rays
rather well, one \halpha\ flare (marked F$_{\rm H\alpha}$) occurs when
the X-rays are low and declining; this appears unrelated to X-ray
behavior.  The continuum apparently tracks the X-ray behavior less
well than \halpha.

\begin{figure}
\includegraphics[angle=90,scale=0.35]{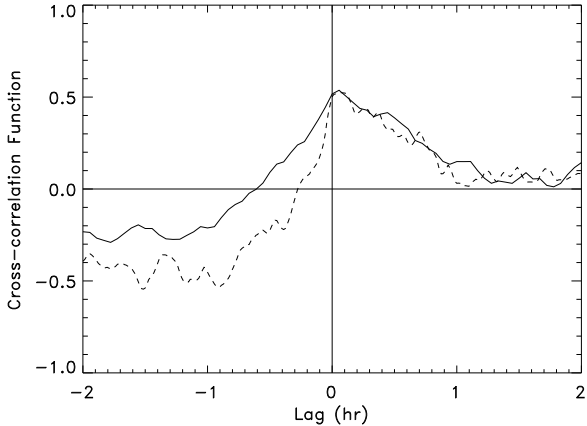}
\caption{X-ray vs.\ \Halpha\ cross-correlation functions.  The solid
  line is derived from the first (WHT) segment.  The dashed line
  corresponds to the Gemini segment, terminated after flare F$_4$ to
  avoid contamination by the \halpha\ flare F$_{\rm H\alpha}$.
  Positive lags correspond to the optical lagging the X-rays.}
\label{CCFFig}
\end{figure}

To measure the lag between the X-ray and \Halpha\ lightcurves we
calculated interpolation cross-correlation functions (CCFs;
\citealt{Gaskell:1987a}; \citealt{White:1994a}).  These are shown in
Fig.~\ref{CCFFig}.  The WHT and Gemini data were used unbinned, and
cross-correlated against X-ray lightcurves with 200\,s and 50\,s
time-resolution respectively.  These CCFs exhibit a very similar
structure to the line vs.\ continuum CCFs presented by
\citet{Hynes:2002a}.  There is clearly no large lag in the line
response to within a few hundred seconds.  Viscous, thermal, and even
dynamical timescales in the line-formation region are likely to be
greater than this, so coupling on these timescales is inconsistent
with the observations.  A lag corresponding to the light travel time
across the disk ($\la40\,s$) is possible, and indeed the data suggest
a positive lag larger than this.  We do not claim detection of a
non-zero lag without more detailed examination of the data, however,
and defer this to a later work.  If the lag is larger than expected
from light travel times, then one explanation might be a finite
reprocessing time, as this might be rather large for the cool
atmospheres expected in the disk in quiescence
(\citealt{Cominsky:1987a}; \citealt{McGowan:2003a}).

\section{Line Profiles}

\begin{figure}
\includegraphics[angle=90,scale=0.35]{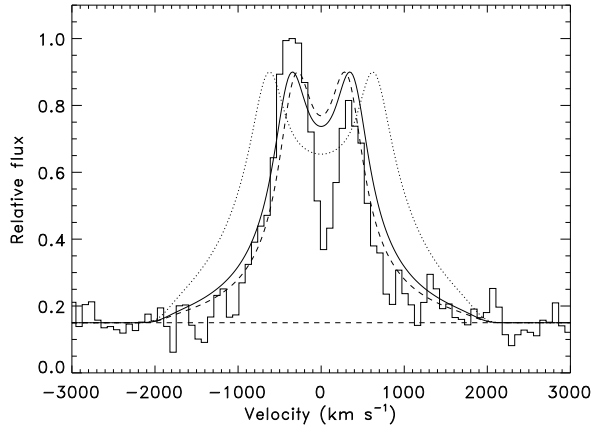}
\caption{Optical spectrum near \halpha\ of F$_1$.  The
histogram shows the data.  Note that the continuum flare spectrum has
not been subtracted; this is approximated by the dashed line.  The
smooth lines show representative optically thin line profiles.  The
solid line corresponds to an outer edge at $R_{\rm circ}$; dashed
lines correspond to a tidally truncated disk and
dotted lines to a disk extending to $R_{\rm circ}/3$.}
\label{ProfileFig}
\end{figure}

We show in Fig.~\ref{ProfileFig} a section of the spectrum of flare
F$_1$, calculated as the difference between the spectra during a
trough at 21.5\,hrs and the peak at 22.5\,hrs.  The \halpha\ line
clearly exhibits double-peaked enhancements during the flare, similar
to the difference profiles obtained by \citet{Hynes:2002a}.  This
indicates that the line response is dominated by the accretion disk
rather than by the companion star or stream-impact point.  We have
shown a representative optically thin line profile for comparison,
smoothed to match the spectral resolution.  The only characteristics
we have attempted to reproduce are the flux and separation of the
peaks.  The other parameters assumed are an inner disk radius
$10^4$\,R$_{\rm sch}$, black hole mass 12\,M$_{\odot}$, and
inclination 56$^{\circ}$ \citep{Shahbaz:1994a}.  We assume the
emission line surface brightness varies as $R^{-1.7}$
\citep{Horne:1986a}.  The peak separation is well fitted assuming a
disk with outer edge at the circularization radius, estimated to be
$R_{\rm circ}=(9.2\pm0.4)\times10^{11}$\,cm assuming parameters from
\citet{Casares:1994a} and \citet{Shahbaz:1994a}.  A significantly
smaller outer disk radius, as shown in Fig.~\ref{ProfileFig}, is not
consistent with the data, although it is possible for the illuminated
region to extend to the tidal truncation radius, $\sim
1.3\times10^{12}$\,cm.  This profile is clearly inadequate in other
ways; it does not reproduce the wings or central minimum well, and the
asymmetry of the peaks is unaccounted for.  The key result for our
purposes, the outer radius of the emitting region, is largely
insensitive to modifications to reproduce the shapes of the wings and
the core, and the asymmetry, and so is a robust result for this flare.
The other flares are shorter and/or weaker, so the flare line profile
is less well constrained.

Material outside of $R_{\rm circ}$ is expected to have a dynamical
timescale of $\ga6$\,hrs.  This is much longer than most of the events
in the lightcurve, in particular the early rise to F$_1$ from which
the flare line profile was derived, and is also much longer than any
lag between the X-ray and \Halpha\ lightcurves (Fig.~\ref{CCFFig}).
The only plausible way to couple the whole line profile to the X-ray
variations is therefore through irradiation, as the light travel time
to $R_{\rm circ}$, $\sim30$\,s, is easily short enough.  We note that
the dynamical timescale at $R_{\rm circ}$ is actually comparable to
the 6\,hr quasi-periodicity reported in both photometry and line
behavior by \citet{Casares:1993a} and \citet{Pavlenko:1996a}.

\section{Flare Energetics}

\begin{figure}
\includegraphics[angle=90,scale=0.35]{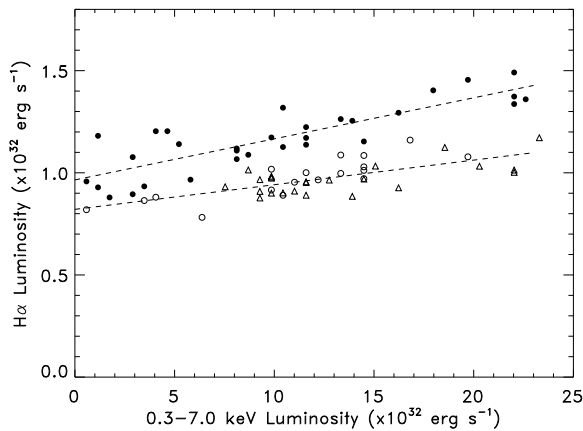}
\caption{Relationship between X-ray and optical luminosities.
  Conversion to luminosity assumes a distance of 3.5\,kpc, constant
  X-ray spectral shape, absorption column $8\times10^{21}$\,cm$^{-2}$
  (estimated from a power-law fit to the 0.3--7.0\,keV spectrum) and
  optical extinction $A_V=4.0$.  The X-ray emission is assumed to be
  isotropic.  \Halpha\ emission has been corrected assuming it
  originates from a disk at an inclination of 56$^{\circ}$.  Solid
  points correspond to the interval D$_1$--D$_2$; open symbols
  correspond to D$_2$ to the end of F$_4$.  The period after F$_4$ was
  excluded due to the presence of the \halpha\ flare F$_{\rm
  H\alpha}$.  Circles indicate WHT data, triangles are from Gemini.
  Dashed lines indicate linear fits.}
  \label{CorrelationFig}
\end{figure}

Fig.~\ref{CorrelationFig} shows the relationship between X-ray and
\Halpha\ luminosities.  There is no evidence for an uncorrelated X-ray
component; the correlation extends to almost zero X-ray flux and the
lightcurves show no X-ray features which are not reproduced by
\Halpha\ (except possibly at the very end of the lightcurve).  There
is, however, an approximately constant component to the \halpha\
emission, which varies only slowly.

From the linear fits shown in Fig.~\ref{CorrelationFig}, the variable
component of \halpha\ luminosity corresponds to 2.0\,\%\ of the X-ray
luminosity in the first segment and 1.2\,\%\ in the second.  Neither
of the quantities compared are bolometric luminosities.  The observed
X-ray luminosity (assumed to be isotropic) is a lower limit on the
bolometric irradiating luminosity, which also includes EUV and
$\gamma$-ray emission.  \halpha\ provides a lower limit on the
reprocessed luminosity.  More detailed modeling will be needed to
estimate these bolometric corrections.  If we consider only
irradiation that can ionize neutral hydrogen (i.e.\ above 13.6\,eV),
and below 100\.keV then this is unlikely to exceed the X-rays by more
than a factor of a few; for example, it is about a factor of three for
a pure power-law spectrum (photon index $\Gamma=1.8$) and a factor of
five for model 1 of \citet{Narayan:1997a}.  Such a low-energy cut-off
is somewhat arbitrary, but is also motivated by the large uncertainty
concerning optical synchrotron emission in the models of
\citet{Narayan:1997a}.  This contributes most of the truly bolometric
luminosity, but is much weaker in more recent models (e.g.\
\citealt{Quataert:1999a}; \citealt{Ball:2001a}).  \halpha\ will not
exceed 20--30\,\%\ of the reprocessed luminosity, where the limit
corresponds to Case B recombination \citep{Osterbrock:1989a}; it is
likely to be substantially less than this.  Thus the reprocessed
fraction is likely to be at least a few per cent, although this is is
not a solid, model-independent constraint.  For a thin disk and
isotropic irradiation, the fraction intercepted is approximately
$H/R$, so the lower limit is plausible for a central compact X-ray
source irradiating a thin disk ($H/R\ga0.02$).  However, there is also
a significant component of the optical continuum which is correlated
with X-rays (Figs.~\ref{MWLCFig}~\&~\ref{ProfileFig}).  If this also
originates in reprocessed X-rays then the reprocessed fraction would
be larger, as the optical continuum flux exceeds that in \Halpha\ by a
factor much larger than plausible bolometric corrections to the
irradiating flux.  This case would then favor an elevated or
vertically extended X-ray emission geometry which allows more
efficient illumination of the disk.  The variable optical continuum
component could alternatively be dominated by synchrotron emission
(e.g., \citealt{Kanbach:2001a}; \citealt{Hynes:2003a}).

\section{Conclusions}
We have established that optical and X-ray variations in \target\ in
quiescence are fairly well correlated.  All X-ray variability
(accounting for essentially all of the observed X-ray flux) is
mirrored well by \halpha, and to a lesser extent by the optical
continuum.  There is clearly another component of \halpha\ emission,
which exhibits rarer, or less pronounced variations, but is not
completely constant.  The correlated \Halpha\ component exhibits
double-peaked line profiles indicating emission from a disk.  The peak
separation implies that the outer edge of the emitting region is at or
outside the circularization radius.  The timescales of the flares,
significantly less than the dynamical timescale at the circularization
radius, suggest that the X-ray--\halpha\ connection is mediated by
irradiation of the accretion disk.  The correlated \halpha\ has a
luminosity of approximately 1--2\,\%\ of the 0.3--7.0\,keV X-ray
luminosity, which is consistent with an irradiation model.  Our
results therefore demonstrate that X-ray/EUV irradiation has a
measurable effect even in quiescent BHXRTs, and that optical
observations can be used to perform an indirect study of X-ray (i.e.,
inner disk) variability, at least for \target.
\acknowledgments
RIH is supported by NASA through Hubble Fellowship grant
\#HF-01150.01-A awarded by STScI, which is operated by AURA, for NASA,
under contract NAS 5-26555.  Chandra observations were supported by
NASA grant GO3-4044X.  The WHT is operated on La Palma by the ING in
the Spanish Observatorio del Roque de los Muchachos of the Instituto
de Astrof\'\i{}sica de Canarias.  The Gemini Observatory is operated
by AURA, under a cooperative agreement with the NSF on behalf of the
Gemini partnership: NSF (United States), PPARC (United Kingdom), NRC
(Canada), CONICYT (Chile), ARC (Australia), CNPq (Brazil) and CONICET
(Argentina).  This work has also made use of the NASA ADS Abstract
Service.

\end{document}